\documentclass[11pt]{cernrep}
\usepackage{epsfig, amsmath}

\begin{document}

\newcommand{\fs}[1]{/\!\!\!#1}

\title{Bound states in the LFD Yukawa model}
\author{M.~van~Iersel and B.L.G.~Bakker}
\institute{Vrije Universiteit, Amsterdam, the Netherlands}

\maketitle

\begin{abstract}
Our purpose is to calculate relativistic bound states in a quantum field
theoretical approach. We work in the Yukawa model and first calculate the
bound-state equation in the ladder approximation. We discuss why this is not a
complete treatment and what possibilities there are to extend this equation.
\end{abstract}

\section{Introduction}

Light Front Dynamics (LFD) is an approach which is very well suited to the
problem of calculating bound states in a quantum field theoretical approach (see
\cite{ben}, \cite{bpp}). In previous work \cite{bvip} we have calculated the
bound states in the LFD approach using a scalar model and compared them with
calculations done in the instant form. Here we move to a more realistic model
where spin is included for the constituent particles. Work on this subject has
already been done by others, see e.g. \cite{ghpsw,kcm,bhm,sfcs}. One aspect that
has not gotten that much attention in the literature is the inclusion of
instantaneous terms in a bound-state equation.

In section 2 we detail the model and derive the bound-state equation
in the ladder approximation. Section 3 deals with some differences between
propagating and instantaneous terms in a bound-state equation. In the last
section we give some methods that may be useful in generating an equation which
contains the instantaneous as well as propagating terms to all orders.

\section{Ladder approximation in the Yukawa model}

The model we are using is the Yukawa model, i.e., we have two particles of spin
$s = 1/2$ and equal mass $m$ which exchange a scalar particle of mass $\mu$. The
Langrangian for this system is given by:
\begin{eqnarray}
\mathcal{L} &=& \frac{i}{2} \bar{\Psi} \gamma_{\mu} \left( \partial^{\mu} \Psi
\right) - \frac{i}{2} \left( \partial^{\mu} \bar{\Psi} \right) \gamma_{\mu} \Psi
- m \bar{\Psi}\Psi + \frac{1}{2} \partial_{\mu} \phi \partial^{\mu} \phi -
\frac{\mu^{2}}{2} \phi^{2} - g\phi\bar{\Psi}\Psi \, ,
\label{eqn1}
\end{eqnarray}
where $g$ is the coupling constant, $\phi$ the scalar field of the exchanged
particle and $\Psi$ the field of the constituent particles. Restricting the Fock
space to the two- and three-particle sector only and not including any self
energy terms, it is possible to derive the bound-state equation from this
Lagrangian. Before deriving this equation one has to realize that spin-$1/2$
constituents have some peculiarities, which result in the fact that we need to
express our equation in terms of the independent degrees of freedom: $\Psi_{+}$,
$\bar{\Psi}_{+}$ and $\phi$. The free field expansion for the fermionic field,
$\Psi_{+}$, involve the 'good' spinors $u_{+}(p,s)$ (or $v_{+}(p,s)$). These 'good'
spinors can be related to the more commonly used spinors $u(p,s)$ (or $v(p,s)$)
via the projection operator $\Lambda_{+}$; $u_{+}(p,s) = \Lambda_{+} u(p,s)$. The
spinors we use were defined before by Kogut and Soper \cite{ks}.

\begin{figure}[h]
\begin{center}
\epsfig{figure=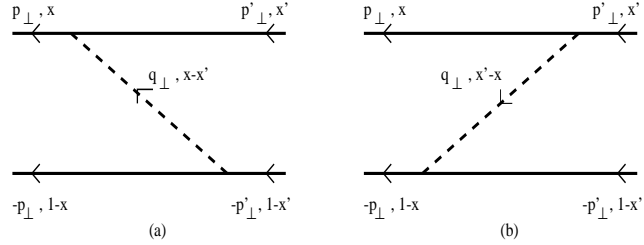,width=85mm,height=32mm,angle=0}
\caption{Time-ordered diagrams in the ladder approximation for the one-particle
exchange.}
\label{todiagram}
\end{center}
\end{figure}

Along these lines we find the equation for the bound state of a fermion and an
anti-fermion
\begin{eqnarray}
\left[ M^{2} - \frac{\vec{p}_{\perp}^{\,2} + m^{2}}{x(1-x)} \right]
\psi(\vec{p}_{\perp},x) &=& \frac{g^{2}}{2(2\pi)^{3}}
\sum_{\sigma_{1},\sigma_{2}} \int \left[ \mathrm{d}^{3}p' \right]
\mathcal{K}(\vec{p}_{\perp},x;\vec{p}_{\perp}^{\,\prime},x')
\psi(\vec{p}_{\perp}^{\,\prime},x') \,.
\label{eqn2}
\end{eqnarray}
Here is $M$ the total mass of the system, $\mathcal{K}(\vec{p}_{\perp},x;
\vec{p}_{\perp}^{\,\prime},x')$ is the kernel of the equation and the integration
element is $[\mathrm{d}^{3}p] = \mathrm{d}^{2}\vec{p}_{\perp} \mathrm{d}x$. The
expression for the kernel is
\begin{eqnarray}
\mathcal{K}(\vec{p}_{\perp},x;\vec{p}_{\perp}^{\,\prime},x') &=&
\frac{1}{\Phi(x,x')} \left( \frac{\theta(x-x')}{2(x-x')D_{a}} \langle s_{1}
\sigma_{1} | s_{2} \sigma_{2} \rangle + \frac{\theta(x'-x)}{2(x'-x)D_{b}} \langle
s_{1} \sigma_{1} | s_{2} \sigma_{2} \rangle \right) \,,
\label{eqn3}
\end{eqnarray}
where the phase-space factor $\Phi(x,x')$ is given by
\begin{eqnarray}
\Phi(x,x') &=& \sqrt{x(1-x)x'(1-x')} \,,
\label{eqn4}
\end{eqnarray}
$D_{a}$ and $D_{b}$ are the energy denominators corresponding to the two
time-orderings
\begin{eqnarray}
D_{a} &=& M^{2} - \frac{\vec{p}_{\perp}^{\,\prime\,2} + m^{2}}{x'} -
\frac{\vec{p}_{\perp}^{\,2} + m^{2}}{1-x} - \frac{(\vec{p}_{\perp} -
\vec{p}_{\perp}^{\,\prime})^{2} + \mu^{2}}{x-x'} \,, \nonumber\\
D_{b} &=& M^{2} - \frac{\vec{p}_{\perp}^{\,2} + m^{2}}{x} -
\frac{\vec{p}_{\perp}^{\,\prime\,2} + m^{2}}{1-x'} -
\frac{(\vec{p}_{\perp}^{\,\prime} - \vec{p}_{\perp})^{2} + \mu^{2}}{x'-x} \,,
\label{eqn5}
\end{eqnarray}
and the spin matrix elements are given by
\begin{eqnarray}
\langle s_{1} \sigma_{1} | s_{2} \sigma_{2} \rangle &=&
\bar{u}(\vec{p}^{\,\prime}_{\perp},x',\sigma_{1}) u(\vec{p}_{\perp},x,s_{1})
\bar{v}(-\vec{p}^{\,\prime}_{\perp},1-x',\sigma_{2})
v(-\vec{p}_{\perp},1-x,s_{2}) \,.
\label{eqn6}
\end{eqnarray}
The diagrams corresponding to this bound-state equation are shown in
Fig.~\ref{todiagram}.

One has to realize that only the propagating diagrams are included in the ladder
approximation. Instantaneous terms are not included and we have to add them to
our equation to get the complete kernel. The fact that these instantaneous terms
are neglected makes it impossible to relate the bound-state equation,
Eq.~(\ref{eqn2}), to the Bethe-Salpeter equation, since in the latter one the
instantaneous terms are taken into account. In perturbation theory the
instantaneous terms were included by e.g. Ligterink and Bakker \cite{lb} and in
the bound-state equation they were included by Sales et al.\cite{sfcs}.

\section{Propagating and instantaneous terms}

As is shown in \cite{lb} it is possible to split the full propagator as follows,
\begin{eqnarray}
\frac{\fs{p} + m}{p^{2} - m^{2} + i\varepsilon} &=& \sum_{\sigma} \frac{u(p,\sigma)
\otimes \bar{u}(p,\sigma)}{p^{2} - m^{2} + i\varepsilon} + \frac{\gamma^{+}}{2p^{+}} \,,
\label{eqn7}
\end{eqnarray}
where the first term is the on-shell, propagating part and the second term is the
instantaneous term.

\begin{figure}[h]
\begin{center}
\epsfig{figure=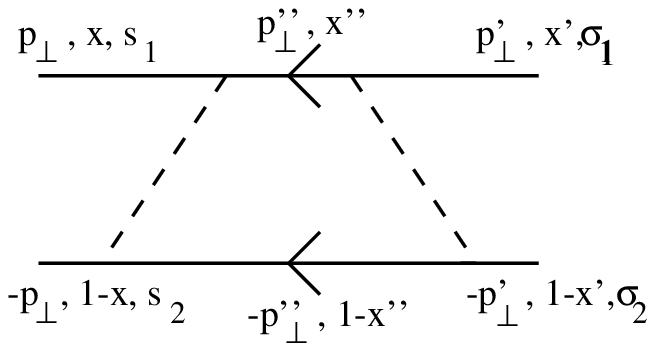, height=32mm, width=55mm, angle=0}
\hspace{5mm}
\epsfig{figure=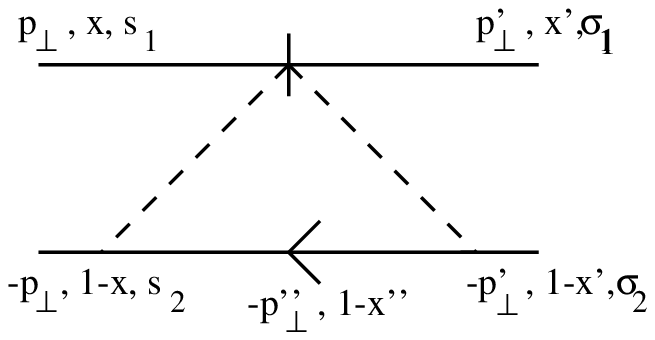, height=32mm, width=55mm, angle=0}
\caption{Two time-ordered diagrams of the covariant box diagram, one containing
only propagating terms (left) and one containing also an instantaneous terms
(right).}
\label{boxpropinst} 
\end{center}
\end{figure}

As an example to illustrate the differences between the propagating and
instantaneous part of the propagator, we will look at one of the time-ordered
box diagrams shown in Fig.~\ref{boxpropinst}. The diagram on the left contains only
propagating terms, the diagram on the right contains also an instantaneous
interaction.
The amplitudes for these diagrams are denoted by
\begin{eqnarray}
\frac{g^{4}}{(2\pi)^{6}} \int \left[ \mathrm{d}^{3} p'' \right]
\mathcal{K}_{2}^{\alpha}(\vec{p}_{\perp},x; \vec{p}_{\perp}^{\,\prime},x';
\vec{p}_{\perp}^{\,\prime\prime},x'') \psi(\vec{p}_{\perp}^{\,\prime},x') \,,
\label{eqn8}
\end{eqnarray}
where $\alpha = P$ or $I$, denoting whether we are dealing with the propagating
or instantaneous interaction. The kernels in both cases are given by
\begin{eqnarray}
\mathcal{K}_{2}^{P} &=& \frac{\theta(x''-x)\theta(x''-x')}{\Phi^{P} D_{a} D_{b}
D_{c}} \bar{u}(\vec{p}^{\,\prime}_{\perp},x',\sigma_{1}) \Lambda
u(\vec{p}_{\perp},x,s_{1}) \bar{v}(-\vec{p}^{\,\prime}_{\perp},1-x',\sigma_{2})
\Lambda v(-\vec{p}_{\perp},1-x,s_{2}) \,, \nonumber\\
\mathcal{K}_{2}^{I} &=& \frac{\theta(x''-x)\theta(x''-x')}{\Phi^{I} D_{a} D_{c}}
\bar{u}(\vec{p}^{\,\prime}_{\perp},x',\sigma_{1}) \gamma^{+} u(\vec{p}_{\perp},x,s_{1})
\bar{v}(-\vec{p}^{\,\prime}_{\perp},1-x',\sigma_{2}) \Lambda
v(-\vec{p}_{\perp},1-x,s_{2}) \,,
\label{eqn9}
\end{eqnarray}
where $\Lambda = \sum_{\sigma} u(p,\sigma) \otimes \bar{u}(p,\sigma)$ and
corresponds to the propagating fermion and $\gamma^{+}$ to the instantaneous
propagator. The energy denominators $D_{i}$ are given by
\begin{eqnarray}
D_{a} &=& M^{2} - \frac{\vec{p}_{\perp}^{\,2}+m^{2}}{x} -
\frac{\vec{p}_{\perp}^{\,\prime\prime\,2}+m^{2}}{1-x''} -
\frac{(\vec{p}_{\perp}^{\,\prime\prime} - \vec{p}_{\perp})^{2} +
\mu^{2}}{x''-x} \,, \nonumber\\
D_{b} &=& M^{2} - \frac{\vec{p}_{\perp}^{\,\prime\prime\,2}+m^{2}}{x''} -
\frac{\vec{p}_{\perp}^{\,\prime\prime\,2}+m^{2}}{1-x''} \,, \nonumber\\
D_{c} &=& M^{2} - \frac{\vec{p}_{\perp}^{\,\prime\,2}+m^{2}}{x'} -
\frac{\vec{p}_{\perp}^{\,\prime\prime\,2}+m^{2}}{1-x''} -
\frac{(\vec{p}_{\perp}^{\,\prime\prime}-\vec{p}_{\perp}^{\,\prime})^{2}+\mu^{2}}{x''-x'}
\,,
\label{eqn10}
\end{eqnarray}
and the phase-space factors $\Phi^{\alpha}$ are given by
\begin{eqnarray}
\Phi^{P} &=& 4 \sqrt{x(1-x)x'(1-x')} x'' (1-x'') (x''-x') (x''-x) \,, \nonumber\\
\Phi^{I} &=& 4 \sqrt{x(1-x)x'(1-x')} (1-x'') (x''-x') (x''-x) \,.
\label{eqn11}
\end{eqnarray}
The differences in the expressions for these diagrams lie in the definition of the
phase-space factor, the number of energy denominators which are present in the
kernel and of course the spin matrix elements.
Looking at the expressions for both diagrams, we see that they both have an
ultraviolet logarithmic divergence.

In \cite{lb} it was shown that in order to get agreement between a covariant and
a LFD calculation on needs to add diagrams containing an instantaneous fermion to
ones that contain only propagating fermions. This way a {\textit{blink}} is
constructed and the divergences of both diagrams appear to cancel. The two
diagrams in Fig.~\ref{boxpropinst} can be added this way, constructing a
{\textit{blink}} on the upper line.

Using this {\textit{blink}} mechanism, we are able to include the instantaneous
diagrams in the bound-state equation. Unfortunately, we have to add them by hand.
Which means that we will get an equation which is correct up to order $g^{2}$, $g^{4}$,
etc. Using this formalism, it seems impossible to generate an equation in such a way
that we include the instantaneous terms and get the equation correct to all orders.

\section{Other methods}

\begin{figure}[h]
\begin{center}
\epsfig{figure=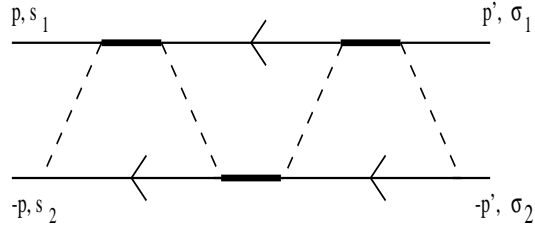,width=70mm,height=30mm,angle=0}
\caption{Higher order diagram with {\textit{blinks}} on both lines.}
\label{doubleblink}
\end{center}
\end{figure}

One possibility to generate such an equation is given in \cite{sfcs}. Here the
light-front equation is split into two equations
\begin{eqnarray}
T(K) &=& W(K) + W(K) \tilde{G}_{0}(K) T(K) \nonumber\\
W(K) &=& V(K) + V(K) \left[ G_{0}^{F}(K) - \tilde{G}_{0}(K) \right] W(K) \,,
\label{eqn12}
\end{eqnarray}
which generate the bound-state equation. Here $T(K)$ is the transition matrix,
$V(K)$ and $W(K)$ are driving terms and $G_{0}^{F}(K)$ and $\tilde{G}_{0}(K)$ are
Green's functions. The question is whether one can generate all diagrams with
these two equations. We have to look whether it is possible to generate for example
a diagram with {\textit{blinks}} on both legs as is shown in Fig.~\ref{doubleblink}.

An other possible way to generate an equation which has the instantaneous terms
included to all orders, was introduced by Pang and Ji \cite{pj}. The full
propagator is split as follows
\begin{eqnarray}
\frac{\fs{p}+m}{p^{2}-m^{2}+i\varepsilon} &=& \sum_{\lambda} \frac{u_{\lambda}(p)
\bar{u}_{\lambda}(p)}{p^{2}-m^{2}+i\varepsilon} + \frac{w_{\lambda}(p)
\bar{w}_{\lambda}(p)}{p^{2}-m^{2}+i\varepsilon} \,,
\label{eqn13}
\end{eqnarray}
where the $w$-spinors are given by (note that the representation used in \cite{pj}
is different from the one we use)
\begin{eqnarray}
w_{\uparrow}(p) = \frac{\varepsilon(p)}{\sqrt{2p^{+}}} \left( \begin{array}{c}
0 \\ 1 \\ 0 \\ 1 \end{array} \right) \,, \hspace{15mm}
w_{\downarrow}(p) = \frac{\varepsilon(p)}{\sqrt{2p^{+}}} \left( \begin{array}{c}
-1 \\ 0 \\ 1 \\ 0 \end{array} \right) \,,
\label{eqn14}
\end{eqnarray}
where $\varepsilon^{2}(p) = p^{2} - m^{2}$. Further investigation is needed to
see whether it is possible to generate an equation which has all propagating
and instantaneous terms included.

\end{document}